# AI as a Sparring Partner – an HCAI Approach to Promote Human Capabilities

Thomas Herrmann [0000-0002-9270-4501]

Ruhr-University Bochum, Institut für Arbeitswissenschaft, Bochum, Germany
Thomas.Herrmann@ruhr-uni-bochum.de

**Abstract:**
A systematic literature reveals that the role of AI as a sparring partner (SP) is often proposed but not systematically analyzed or defined. We propose the definition: An AI sparring partner (AI SP) interacts with users in a combination of a cooperative as well as a challenging, competitive mode, where AI is selected, customized or self-adapting to meet a level of skills that neither over- nor under-challenges the users. They can have the experience that they *become* better or *are* better than AI. AI as a SP can support creativity, extend viewpoints or foster learning and critical thinking either towards ideas and decisions or towards the AI itself. Sparring with AI can either be explicit, e.g. when AI simulates certain roles, or implicit, when AI is used to find out whether or how to perform better.

## 1. Introduction

Human-Centered AI (HCAI) not only envisions that humans and AI are better together than individually, but also that both can improve through mutual interaction [1]. In this context, AI usage aims at keeping the human in the loop through taking over roles such as an assistant, a ghost writer, a collaborator and teammate, a facilitator, a teacher, as power tool etc. These roles can be relevant in many cases, e.g. when people are only looking for support and do not intend to exert themselves at the highest level of their own skills and abilities. However, there may be some areas where they seek to improve their competencies and skills and be experts that are better than others. We assume that there are people who occasionally or systematically interact with AI in such a way where they believe that they can do better than AI or at least become better. For these use cases, a further role is relevant: using AI as sparring partner (SP) for the purpose of learning.

Academic definitions for SP are hard to find. There are guidelines for practitioners that outline characteristics of a suitable SP (see Section 2). In the context of sport, sparring partners are selected that are similarly strong like a trainee or can adapt their strength.

A sparring partner is a friendly opponent who challenges a person's skills in order to improve them - either selectively or in their entirety. An SP constantly pushes his trainees to their limits but avoids harming them. Accordingly, sparring partners have no concerns that the trainee will outperform them. And a SP helps learners to understand the limits of their capabilities.

The concept of a sparring partner for competitive sports can be transferred to other areas such as manager training or preparing someone for intellectual debates. With an AI sparring partner, humans can develop further by critically engaging with the AI, questioning its results or trying to identify its limitations. Although some scholars refer to the role of AI, in particular generative AI, as a sparring partner [2] [3], the literature on using AI for support of learning and critical thinking does not systematically refer to this concept. Thus, the goal of this paper is to close this gap.

In what follows, we want to explore the potential of AI as a sparring partner for learning and, to this end, investigate the following questions:

- What use cases and characteristics are associated with the concept of AI as a sparring partner in the literature?
- What do the usage patterns of an AI SP look like and how are they socio-technically embedded in the context of organizational practices?

## 2. Background

It is hard to find scientific definitions for "Sparring Partner although this term is multiply used in literature. The Meriam Webster dictionary [4, p. 2183] offers:

> "Sparring: … "1 : scientific boxing
> 2 : a skirmishing for advantage; esp : ARGUMENT, DISPUTE"
> sparring partner … 1 : a boxer's companion for practice in sparring during training
> 2 : one's mate in amicable wrangling or debating: a protagonist with whom to sharpen one's wits in argument."

Wikipedia mainly refers to sparring [5] within combat sport and considers "argumentative debate" to be an extension. In sport, it is organized in a way that prevents injuries. With regard to the organization of sparring, a difference is made depending on whether the sparring takes place between people who know each other and are friends, or with a stranger. Sparring generally differs from competitions, as the aim of sparring is usually to train the participants.

A web site about sparring reveals [6]:

"A sparring partner is someone who pushes you out of your comfort zone, challenges you, and enables you to learn new things. They are essential to our growth because they help us become better people.

**What Is a Sparring Partner, and What Do They Do?**

- A sparring partner is a person trained in the same activity as you who can help you improve your skills.

- In the context of mental health, a sparring partner is someone who challenges you to grow and overcome obstacles.
- Sparring partners are essential because they:
- Help us feel more comfortable around other people
- Improve our communication skills
- Act as healthy competition
- Inspire us to do better.
- A good sparring partner can push you to be better than you are at any given time. A sparring partner will often have similar strengths and weaknesses as the person they are training with. This allows them to push each other to improve their skills and stay sharp."

We consider using AI as a sparring partner as an approach that is typical for human centered AI, since it refers to helping people develop their skills and strengths as well as their creativity [7]. For HCAI, the term sparring relates more to cognitive tasks such as debating and critical thinking. The discussion on how AI can be employed for learning has been given new momentum by the availability of generative AI, such as ChatGPT [8], [9], [10]. One goal of AI based learning is the promotion of critical thinking (CT) [11]: prompting human actors in the context of AI usage is seen as a suitable form of interaction to initiate critical thinking.

According to [12], CT can be understood as taking responsibility of one's own mind by applying logical and reflective thinking and knowledge about how to make valid inferences and abstractions. It is based on the acceptance of the general need for evidence and the reasons for its different levels of accuracy. In the context of using AI, it is relevant that CT includes identifying appropriate sources of information, analyzing its credibility, and reflecting on its consistency with other sources of knowledge [12].

CT is not necessarily meant as a critical attitude towards the performance and the outcome of AI. Kasneci et al. [8] consider the promotion CT as a way where students learn early about the potential societal biases and risks of AI. Mollick & Mollick [13] propose learning processes where students understand that AI's first solution for a task, such as writing an essay, has always shortcomings and that students are instructed to seek step by step for improvement by applying adequate prompts.

Bjelleand et al. [14] suggest a relationship between critical thinking and the usage of AI as a sparring partner. In general, the promotion CT is considered a learning concept for higher education. Bucher & Schenk [3] emphasize that AI as a SP can also be applied in the context of workplace learning. Apparently, seeing AI as a sparring partner can include both: that AI can be challenged by the user or that AI can challenge the human side. The latter aspect was discussed early on with the concept of critiquing systems [15].

An initial review of the basic literature on the relationship between AI usage and HCAI-oriented learning support reveals that the concept of AI as SP is related to learning and critical thinking, but that there is a lack of systematic analysis of the various ways in which AI can be employed as SP and the characteristics that such a SP should have.

## 3. Method

The research questions are addressed by means of a narrative and a systematic literature search. The narrative literature analysis considers foundational papers [8], [9], [10] dealing with the possibilities and challenges of AI, especially regarding the advancement of human abilities through learning with AI. Based on the author's experience, additional approaches to skills development and reflection as well as human-centered AI are considered and discussed in the background section.

The systematic literature analysis identified 650 papers using the keywords "Artificial Intelligence," "Sparring Partner," and "learning" from 2015 to 2024. The number of papers addressing the topic in 2024 has grown rapidly (approximately 70 new papers from mid-August 2024 to early October, and a further 60 articles by mid-January 2025). We found that the term sparring partner can also refer to situations where AI is trained by a SP. We excluded these papers as well as those papers that refer to a person serving as a sparring partner. Table 1 describes how we sorted out the relevant works.

**Table 1:** Sorting out relevant literature.

| Step | Filtering strategy | Papers |
|---|---|---|
| 1 | “artificial intelligence” "sparring partner" learning from 2015-2024 | **650** |
| 2 | Cited less than one time a year | 353 |
| 3 | Papers that have been inspected | **297** |
| 4 | Papers that were irrelevant for certain criteria | 255 |
| 5 | Papers relating SP to AI just occasionally (1-3 times) | **25** |
| 6 | papers providing an in-depth consideration or f the relation between SP and AI | **16** |

The criteria for excluding papers as irrelevant were:

- SP refers to persons mentioned in the acknowledgements
- SP refers to persons who were employed to serve as a SP
- SP refers to training of AI with AI
- The paper just mentions that AI can be a SP, without giving any further indication what this might mean or what examples might look like.
- Paper is not in English

For the succeeding analysis we used the 25 papers with only marginally mentioning SP to find at least other roles of AI that were used as a contrast to SP, or to identify synonyms that are used to circumscribe the concept of AI as a SP (see Tabel 2). We deepened the description of various categories of use cases for AI-based SPs by exploring the remaining 16 papers.

## 4. Findings

Basic results of the literature analysis show that although the concept of the sparring partner is frequently mentioned, it is not systematically examined; in particular, occupational learning in the workplace is widely neglected. The integration of the concept in socio-technical contexts is also not systematically considered.

**Table 2:** Extracted Aspects of AI as a Sparring Partner

| **Roles in contrast to SPs** | **Synonyms for SP** | **Characteristic of SPs** | **What can be done with SPs** |
|---|---|---|---|
| Teacher [16],<br>Teaching assistant [14] | devils advocacy [17] | Challenging:<br>Uncooperative [18]<br>Attacker [19]<br>Outsmarting s.b [20] | For debating:<br>• stimulate discussions [21]<br>• To argue with [22]<br>• provide counterarguments [23] |
| Facilitator [24],<br>Advisor [25],<br>Coach [26] | Pedagogical agent [27] | not necessarily better than the trainee [28] | Simulate discussions [29]; Simulate certain roles [3] |
| Tutor [16] [3] | Learning companion [30] | Implicit teaching aid [20] | Stressing functionality of a system [18]<br>Simulate attacks to train somebody [19] |
| research assistant [31],<br>Expert [24],<br>Analyst [24] | Training partner [28] | Supporting co-evolution [32] | Help become more creative [33]:<br>• Creating stories [34]<br>• For ideation [35]<br>• Idea generation [14], [36] |
| writing collaborator [31],<br>Ghostwriter [16] | co-creation partner [35] | providing social interaction and companionship [26]<br>Being a Peer [24] | Support scientific skills:<br>• generating convincing samples and generating samples that differ from the real distribution [37]<br>• For academic writing [14]<br>• Improve surveys [38]<br>• develop stringent research designs [39] |
| Evaluator [ 22,25],<br>Critical evaluation [14] | | | |
| idea generation [14] | Coach [26] | Being intellectual [31] | encourage and motivate physical and mental exercises [26];<br>Increasing students' interest [40]] |
| Teammate [24],<br>(merged into a human group [41] | Tutor bot [42] | Usable when ever you want [43] | Providing a new viewpoint [44]:<br>• provide interesting ideas and suggestions [16]<br>• For having a second opinion [45]<br>• Reminding things to consider [43] |
| Project office [25],<br>Administrative assistant [31] | | collaborative tool [46] | for critical thinking:<br>• point out flaws in your ideas [23]<br>• "evaluating ideas" [44]<br>• Challenge biases [38]<br>• challenges, refines, and expands their ideas [46] |

It can be seen that in many studies the "sparring partner" is only mentioned once in connection with AI and there is no in-depth analysis of the relationship between AI and sparring. To deepen the relationship between SP and AI (see Table 2), we look at other roles used to make a difference to an AI SP (column 1), synonyms for SP (column2), characteristics and ways of using an SP (columns 3 and 4).

In the first column of Table 2, various roles are presented which, in the authors' view, differ from the role of an AI as a SP. The roles mentioned here can alternate with the use of AI as a sparring partner; for example, a teammate, coach or teacher can temporarily switch to the role of SP. What all these roles have in common is that they keep the human in the loop by enabling close interaction with the AI. Column 2 shows synonyms used to describe the role of a SP. The comparison with column 1) reveals that the concept of AI as SP is not clearly specified. While some authors consider tutors and coaches as synonyms for SP, others see the same roles as something different. All synonyms in column 2 emphasize that SPs help with learning and improving skills. A special case is the characterization as “devil’s advocate”, which focuses on a type of learning that provokes the learner by contradicting their statements or beliefs.

The characteristics of an SP listed in column 3) emphasize the difference between two types of sparring: one can be strictly competitive (attacking, outsmarting), the other more cooperative (being a peer, collaborative tool, teaching aid). While many descriptions of AI implicitly suggest that AI could be a superior challenger, other views propose that AI need not be better than the human counterpart (trainee). Both sides, AI as SP and the trainee, could strive for co-evolution. It becomes clear that AI-related SPs target cognitive tasks and improvements (being intellectual) through a type of social interaction that is constantly available.

Column 4) collects aspects of how AI can be used as an SP. We have formed eight different groups. Debating and stimulating discussion seems to be a category that underlies all other ways of using the SP: an exchange of points of view or arguments takes place. Simulating discussions from the perspective of different roles, such as a customer, a manager, a consultant, etc., represents use cases for which AI needs to be specially designed, configured or prompted. The same applies to the simulation of attacks to which a user should learn to react (an example that comes from the military context). Such simulations can combine learning with testing (stressing) to see if a person is able to deal with certain types of challenges. In general, sparring is not only a method to support learning, but also to evaluate learners – either by a third party or by themselves.

Many papers that mention SP as a possible role for the use of AI emphasize the potential of supporting creativity, especially ideation, to broaden the range of ideas that can be considered by a user. Providing viewpoints, second opinions or further aspects that can be considered is a type of AI contribution that challenges users to check that they are aware of all relevant aspects when working on a task. Some authors suggest that sparring can increase learner motivation and interest. It should be mentioned here that the aspect of gaming – which is obviously an important aspect in the intersection of learning and sparring – is not explored in depth by any of the literature found. One particular area targeted by the papers analyzed is the use of AI as an SP that helps to improve scientific skills through different types of contributions relevant to research.

Another potential use that is widely addressed in the articles we researched is the promotion of critical thinking. This is a focus that is also pointed out in the literature, which generally [11], [13] focuses on the possibilities of learning with AI. Critical thinking can be triggered by AI by evaluating users' ideas, pointing out shortcomings, challenging biases or further aspects to be considered.

It is easy to see that the eight groups listed in column 4 overlap, mostly in the sense that they can support each other. Consequently, these eight groups are not mutually exclusive, but are strongly intertwined. However, it can be pointed out that the use of AI as a SP may require different modes of interaction depending on the use case in focus.

Referring to the literature that elaborates on the relationship between AI and sparring, we can complete the findings presented in Table 2. Further aspects of critical thinking are that people can evaluate their level of skills or knowledge:

- learners can test their knowledge with AI before taking a formal assessment, [42]
- an informative assessment [2] assists students to understand their mistakes or to detect their learning gaps,
- the process of evaluation can be carried out privately, without the judgment from others.

These approaches focus on critiquing or evaluating learners' assumptions about their level of competence. The anonymity of AI as a technical system is seen as an advantage that could reduce learners' anxiety about assessment: Zhu & Deng [47] found that people who are socially anxious are more willing to accept a robotic training partner. This observation underlines the advantage of an SP that is easily and informatively accessible [43].

In the general literature on the use of AI for learning and critical thinking, the importance of a critical attitude towards AI itself is emphasized [8]. However, this aspect is only marginally considered in the literature that views AI as an sparring partner: Herrmann[48] points out that humans may want to be or become better than AI in certain skill niches and therefore use AI as SP to see how they can improve their skills to be better. As a typical scenario in which AI is used as a sparring partner for the continuous development of one's own skills, he describes a scenario in which one first prompts the AI to provide a solution and then tries to see how one can significantly exceed the level of the solution generated by the AI in terms of individuality, creativity, regional characteristics, etc.

That a critical view of AI may be appropriate is expressed in papers that focus on using the SP to enhance creativity. Possible hallucinations of some AI bots are considered bearable and not as a problem as long AI is only used as a SP that helps to expand the range of one's ideas [49]. Using AI as a SP can involve experimentation to generate interesting ideas and suggestions [16], always being aware that not all ideas are good [36], but are delivered faster, and possibly are broader and at least thought-provoking. If creative thinking is understood as a back and forth between divergence (producing many ideas) and convergence (selecting ideas), then critical thinking complements ideation by supporting convergence [50].

There is also little in the papers about processes [51] that describe how AI can be used as SP, particularly for critical thinking. There are few references: Reis et al. [29] suggest that AI be asked to propose topics to be studied and then test learners on those topics. Others [52] favor a back and forth strategy where one first expresses one's own ideas on a particular topic and then asks the system for other ways of thinking or methods to approach that idea. Or you choose an idea that doesn't sound plausible and ask the system to convince you [21]. The basic assumption is that idea generation with AI always alternates with critical reflection. To summarize, critical thinking always involves two sides: a careful reflection on the ideas you have generated yourself or with AI, and a critical reflection of the capabilities of AI itself.

To define the characteristics of an AI SP, it is important to analyze how the SP should behave or be designed. Magrill & Magrill [53] point out that all resources, feedback and assessments provided by the AI SP must be customized to the needs of the trainee. Referring to the context of boxing, they require that a good sparring partner challenges people only as far as they are willing to go, gradually improving their skills. An SP provides individual feedback and can flexibly adapt its communication to the flow of the conversation [3]. Efficiency is therefore achieved through flexibility in order to avoid wasting time. In conclusion, it can be said that the behavior of an AI SP must be within a corridor in which it neither under- nor overstrains the trainee.

These characteristics can be achieved if AI is explicitly designed or prompted to behave like an SP by simulating certain roles. When an AI bot acts as a certain character, e.g. a customer, users appreciate if it is not easily satisfied and can express emotions to appear as a real-life counterpart for the trainee [3]. In this context, it is also important that the SP is able to provide feedback after the role play. This subsequent feedback is important to understand that supporting reflection is also an essential task in connection with sparring; either other people or the AI itself can serve as a reflection partner [54]. This reflection also has different sides, as is the case with critical thinking: it can relate to the content of the ideas, to the skills of the participant or it encourages participants to think critically about the application of AI in problem solving [16].

Apparently, AI can either be configured to play the role of an SP, even by simulating consciousness [55], or it can simply be used as an SP without being "aware" of this role. For example, a chess program does not know that it is outsmarting a person or that it is a teaching tool [20]. In this type of constellation, learners have to draw their own conclusion in order to achieve a learning effect.

## 5. Discussion and Conclusion

### 5.1. Definition of AI as a SP

Although the term "sparring partner" is frequently used in the literature on the roles of AI there is no clear definition. Based on the more general characteristics (Section 2) and the literature analysis we propose the following definition of the role of AI as a sparring partner:

> An AI sparring partner (AI SP) interacts with users in a combination of a co-operative as well as a challenging, competitive mode, where AI is selected, customized or self-adapting to meet a level of skills that neither over- nor under-challenges the users. They can have the experience that they *become* better or *are* better than AI. These learning effects are based on the support provided by AI, which either expands the range of ideas and viewpoints considered (A) or fosters discussion and critical thinking about and reflecting on (B) the quality of ideas or problem solving (B1), the skills and knowledge of the user (B2) or the performance of the AI system itself (B3).

We can basically differentiate between two modes of SP:

- Explicitly, where AI is designed, configured or prompted to be a SP
  a) AI critically revises a learner's or trainee's input to the system and give hints for improvement
  b) AI is prompted to simulate a role during competitive interaction where it flexibly adapts its performance to the trainee's level
- Implicitly, where AI is just used as a SP
  a) The user prompts AI to solve certain problems, considers and interactively refines its output and tries to go beyond the quality of AI output
  b) The learners or trainees do not directly interact with AI but with other persons who are supported to be a SP by employing AI.

These modes as well as the differentiation that is provided in the definition requires different ways of supporting the interaction with AI in the role of an SP.

Overall, encountering an SP helps people to understand what their own strengths are and whether these strengths are a basis to be better than AI or not. An AI SP helps also to understand whether one overestimates one's own strengths, and whether there is still space to improve one's own performance. It might always be the case that with respect to a certain question or problem, AI is better than a human user. However, this does not exclude the possibility that the humas can identify areas where they can add something that leads to an even better result and that the whole process of creating a better result can be accelerated when doing it the next time. In these cases, the role of AI as a SP is appropriate.

### 5.2. Variants of implementing AI as a SP

Since the definition includes various modes of how AI can be used in the role of a sparring partner, we can assume that there are various ways of how such a usage is started and follows a certain process to be completed. Since the literature does not describe these possible processes, we provide a list of suggestions that can be considered as a first overview of the variety of possible solutions:

- One starts with a prompt to let AI solve a certain problem. When the users see the solution offered by the AI, they can compare it to what they themselves have suggested, think about improvements based on their experience, and describe the improved approach to the problem. Possibly this description can be re-entered to see if the AI is able to counter it with something reasonable. Such a dialog can be repeated until the user assumes that no further improvement can be achieved. Finally, the user – possibly together with the AI – can reflect on the question of whether they or the system or both have made the decisive contribution to the final result.

- The user might assume that s/he has already an appropriate idea of solving a problem, enters into a dialogue with the system to describe this solution and can go on prompting the system with questions such as: Does my solution cover all relevant aspects; what are the arguments against this solution; convince me that I am wrong; what could I improve? Depending on the answers the users may change their proposal or may become convinced that they found the best solution. Potentially this belief is discussed with the system as a way of reflecting.
- Similarly, someone who feels challenged to develop a new concept or idea, e.g. for market entry, could describe their designs for the system and ask if they are really new, or what similar concepts already exist, etc.?
- The entire process of creative work could be accompanied by starting with a brainstorming session. You could check whether AI can expand on the brainstorming results that a team has found. This can lead to a back-and-forth process of changing the prompts to intensify the brainstorming. Secondly, the sorting out of ideas or the integration of ideas can be supported by AI and accompanied with an argumentative exchange that helps the participants to see if they perform better than AI in the step-by-step elaboration of their favorite idea.
- Another case where sparring is started is when someone wants to check whether s/he is fit enough to handle a very specific task or take on a multi-faceted job. For this purpose, a person could describe the task or job to be performed and ask the system to challenge or test them. During the test, a negotiation could begin as to whether the test tasks are appropriate. In a subsequent reflection phase, one can try to understand what possible knowledge or gaps need to be overcome before taking on a job - including whether AI can help to close these gaps.
- Another example would be a sparring session where someone wants to find out whether there are tasks that they can solve and that the AI cannot solve or can only partially solve. In this case, the user must prepare a series of tasks where he/she knows how to check the quality of the solution. To demonstrate his/her superiority, the user must indicate the limits of the AI in a dialog with iterative queries. Possibly the system can admit its shortcomings in a subsequent reflection.

This list demonstrates that there are a number of typical questions that are relevant for each use case where AI serves as a sparring partner:

1. **When should an AI-SP be used or the usage be repeated?**
   Exemplary answers might be:
   a) when someone realizes that they need to improve their skills
   b) to test whether they need to get better
   c) when a combination with other disciplines or an extension of viewpoints and ideas is helpful.

2. **How should an AI-based sparring partner be selected or configured?**

When selecting a partner, it must be possible to find one who neither overchallenges nor underchallenges the trainee. The trainee must have the feeling that s/he can be or become better. The SP must either have the same strengths as the trainee or strengths in an area that is relevant for the trainees and complements their expertise. Or the SP can be stronger but must be able to adapt to the trainee's level, e.g. by employing a user model.

3. **When can or should a training with a sparring partner be finished?**
   The users or those members of an organization who initiate the training with a sparring partner can specify benchmarks to be achieved, indicating that the training can be considered complete. Typical examples for benchmarks by which the user realizes that s/he has the level of a SP are:
   a) the user finds tasks that s/he can solve but AI cannot,
   b) the user can predict what AI will provide,
   c) the users can answer all questions or test that an AI sparring partner asks them.

   Reaching such a benchmark can also mean, that another SP with a higher level of capabilities should be found.

With respect to the list of possible processes provided above it also becomes obvious that most use cases for an AI based SP are embedded into a social organization. Consequently, there are examples for different constellations that indicate that AI based SPs are part of a socio-technical system or process, such as:

- It is not only an individual but a team that can be challenged by an AI-based SP; and it can even be the team together with a certain technical infrastructure, e.g. to maintain cyber security or to prevent machinery breakdowns, that seeks testing with AI.
- Someone may assume that s/he has outperformed AI in completing an intellectual task during a sparring session. This assumption of users can be critically evaluated by their teammates who may confirm their impression or request them – e.g. in the case of generative AI – to try out other prompts that could lead to a better AI outcome.
- A person who needs training might not interact directly with an AI SP, but with another person who uses AI to prepare or to run the sparring session.
- AI can be used for a sparring exercise, but the succeeding reflection phase is conducted with other people, an individual or a team.
- The motivation or initiation of a person's sparring session is the task of a team or a manger. The whole sparring procedure is embedded into organizational practices [56]. Since many examples of AI-based sparring are implicit, it is a way of usage that is not mainly based on the design of particular interaction modes, but a type of AI usage that must be encouraged by a person's socio-organizational context.

The quality of socio-technical systems is tightly depending on communication practices [57]. Sparring for improving critical thinking and debating addresses communication in three ways: it is used for sparring, it aims to improve communication, and it uses meta-communication for reflection [58]. Apparently, the design of sparring with AI is not only a matter of research in the field of human computer interaction research but also relevant to the field of computer-supported cooperative work (CSCW).

### 5.3. Further research

The variants listed above in Section 5.2 are of a preliminary character. They can serve to inspire further empirical research that will help to provide more examples of sparring with AI and to better understand those characteristics of an AI-based SP that provide the most benefit to the people and teams involved. For the use cases where AI is designed to simulate certain roles that a trainee needs to interact with, further research has to provide interfaces that help to configure the SP, make the sparring run smoothly, allow the trainee to intervene when needed [59], allow for adjustment of the role-play and conduct phases of reflection, etc. Further research questions should address the interplay an AI-based SP may have with other possible roles of AI and which advantages and disadvantages – on the side of human actors as well as for AI-systems – can be observed in respect to different use cases.

## Acknowledgements

I would like to thank Gerhard Fischer, Isa Jahnke, Sabine Pfeiffer, Nikol Rummel and Reinhard Stolle for their contribution as sparring partners in deepening the approach described here.

## References

[1] D. Dellermann, A. Calma, N. Lipusch, T. Weber, S. Weigel, and P. Ebel, "The future of human-AI collaboration: a taxonomy of design knowledge for hybrid intelligence systems," in Proceedings of the 52nd Hawaii International Conference on System Sciences (HICSS), 2019.

[2] R. J. Krumsvik, "Artificial intelligence in nurse education – a new sparring partner?: GPT-4 capabilities of formative and summative assessment in National Examination in Anatomy, Physiology, and Biochemistry," NJDL, vol. 19, no. 3, pp. 172–186, Oct. 2024, doi: 10.18261/njdl.19.3.5.

[3] A. Bucher and B. Schenk, "When Generative AI Meets Workplace Learning: Creating A Realistic & Motivating Learning Experience With A Generative PCA," 2024.

[4] P. B. Gove and Merriam-Webster, Inc, Eds., Webster's third new international dictionary of the English language, unabridged. Springfield, Mass: Merriam-Webster, 1993.

[5] "Sparring," Wikipedia. Accessed: Jan. 25, 2025. [Online]. Available: https://en.wikipedia.org/wiki/Sparring

[6] A. Betancourt, "Sparring Partners: The Science Behind Why We Need Them and How They Help Our Mental Health," medium.com. Accessed: Jan. 25, 2025. [Online]. Available: https://medium.com/bottomline-conversations/sparring-partners-the-science-behind-why-we-need-them-and-how-they-help-our-mental-health-b1254fbcd8a2

[7] B. Shneiderman, Human-Centered AI. Oxford University Press, 2022.

[8] E. Kasneci et al., “ChatGPT for good? On opportunities and challenges of large language models for education,” Learning and individual differences, vol. 103, p. 102274, 2023.

[9] S. Küchemann et al., “Are large multimodal foundation models all we need? On opportunities and challenges of these models in education,” EdArXiv, 2024.

[10] O. Ozmen Garibay et al., “Six Human-Centered Artificial Intelligence Grand Challenges,” International Journal of Human–Computer Interaction, vol. 39, no. 3, pp. 391–437, Feb. 2023, doi: 10.1080/10447318.2022.2153320.

[11] Y. K. Dwivedi et al., “Opinion Paper: ‘So what if ChatGPT wrote it?’ Multidisciplinary perspectives on opportunities, challenges and implications of generative conversational AI for research, practice and policy,” International Journal of Information Management, vol. 71, p. 102642, Aug. 2023, doi: 10.1016/j.ijinfomgt.2023.102642.

[12] B. Miri, B.-C. David, and Z. Uri, “Purposely Teaching for the Promotion of Higher-order Thinking Skills: A Case of Critical Thinking,” Res Sci Educ, vol. 37, no. 4, pp. 353–369, Aug. 2007, doi: 10.1007/s11165-006-9029-2.

[13] E. R. Mollick and L. Mollick, “New Modes of Learning Enabled by AI Chatbots: Three Methods and Assignments,” SSRN Journal, 2022, doi: 10.2139/ssrn.4300783.

[14] C. Bjelland, K. Ludvigsen, and A. Møgelvang, “Unveiling the impact of AI chatbots on higher education: Insights from students,” presented at the 18th International Technology, Education and Development Conference, Valencia, Spain, Mar. 2024, pp. 1458–1465. doi: 10.21125/inted.2024.0428.

[15] G. Fischer, “Domain-oriented design environments,” Automated Software Engineering, vol. 1, no. 2, pp. 177–203, 1994.

[16] Y. Walter, “Embracing the future of Artificial Intelligence in the classroom: the relevance of AI literacy, prompt engineering, and critical thinking in modern education,” Int J Educ Technol High Educ, vol. 21, no. 1, p. 15, Feb. 2024, doi: 10.1186/s41239-024-00448-3.

[17] A. Claver, “Devil’s advocacy and cyber space. In support of quality assurance and decision making,” Journal of Intelligence History, vol. 20, no. 1, pp. 88–102, Jan. 2021, doi: 10.1080/16161262.2020.1864863.

[18] P. R. Rood, “Scaling Reinforcement Learning Through Feudal Muti-Agent Hierarchy,” 2022.

[19] V. Terziyan, S. Gryshko, and M. Golovianko, “Taxonomy of generative adversarial networks for digital immunity of Industry 4.0 systems,” Procedia Computer Science, vol. 180, pp. 676–685, 2021, doi: 10.1016/j.procs.2021.01.290.

[20] M. McShane and S. Nirenburg, Linguistics for the Age of AI. Mit Press, 2021.

[21] O. P. Singh, “Artificial Intelligence and Psychoanalysis: A New Concept of Research Methodology,” NPRC Journal of Multidisciplinary Research, vol. 1, no. 1, pp. 19–27, 2024.

[22] A. M. Paul, “Extending Biological Intelligence: The Imperative of Thinking Outside Our Brains in a World of Artificial Intelligence,” in Augmented Education in the Global Age, Routledge, 2023, pp. 157–171.

[23] P. Korzynski, G. Mazurek, P. Krzypkowska, and A. Kurasinski, “Artificial intelligence prompt engineering as a new digital competence: Analysis of generative AI technologies such as ChatGPT,” Entrepreneurial Business and Economics Review, vol. 11, no. 3, pp. 25–37, 2023.

[24] D. Siemon, “Elaborating Team Roles for Artificial Intelligence-based Teammates in Human-AI Collaboration,” Group Decis Negot, vol. 31, no. 5, pp. 871–912, Oct. 2022, doi: 10.1007/s10726-022-09792-z.

[25] I. Susha, T. Van Den Broek, A.-F. Van Veenstra, and J. Linåker, “An ecosystem perspective on developing data collaboratives for addressing societal issues: The role of conveners,” Government Information Quarterly, vol. 40, no. 1, p. 101763, Jan. 2023, doi: 10.1016/j.giq.2022.101763.

[26] N. Fronemann, K. Pollmann, and W. Loh, “Should my robot know what’s best for me? Human–robot interaction between user experience and ethical design,” AI & Soc, vol. 37, no. 2, pp. 517–533, Jun. 2022, doi: 10.1007/s00146-021-01210-3.

[27] A. Bucher, B. Schenk, and G. Schwabe, “When Learning Turns To Surveillance – Using Pedagogical Agents in Organizations,” 2023.

[28] S. Choi, H. Kang, N. Kim, and J. Kim, “How Does Artificial Intelligence Improve Human Decision-Making? Evidence from the AI-Powered Go Program.” 2025. [Online]. Available: https://arxiv.org/abs/2310.08704

[29] F. Reis, C. Lenz, M. Gossen, H.-D. Volk, and N. M. Drzeniek, “Practical Applications of Large Language Models for Health Care Professionals and Scientists,” JMIR Med Inform, vol. 12, pp. e58478–e58478, Sep. 2024, doi: 10.2196/58478.

[30] A. S. Gibbons, “What is instructional strategy? Seeking hidden dimensions,” Education Tech Research Dev, vol. 68, no. 6, pp. 2799–2815, Dec. 2020, doi: 10.1007/s11423-020-09820-2.

[31] S. Atlas, “ChatGPT for Higher Education and Professional Development: A Guide to Conversational AI.,” University of Rhode IslandUniversity of Rhode Island, Kingston, 2023. [Online]. Available: https://digitalcommons.uri.edu/cba_facpubs/548

[32] M. Golovianko, S. Gryshko, V. Terziyan, and T. Tuunanen, “Responsible cognitive digital clones as decision-makers: a design science research study,” European Journal of Information Systems, vol. 32, no. 5, pp. 879–901, Sep. 2023, doi: 10.1080/0960085X.2022.2073278.

[33] P. Faraboschi, E. Frachtenberg, P. Laplante, D. Milojicic, and R. Saracco, “Artificial General Intelligence: Humanity’s Downturn or Unlimited Prosperity,” Computer, vol. 56, no. 10, pp. 93–101, Oct. 2023, doi: 10.1109/MC.2023.3297739.

[34] C. Hess and S. Kunz, “‘ChatGPT, Can you Motivate my Learners?’-Co-Creation of Inspiring STEM Lessons with AI Chatbots,” presented at the 16th annual International Conference of Education, Research and Innovation, Seville, Spain, Nov. 2023, pp. 6103–6110. doi: 10.21125/iceri.2023.1522.

[35] D. Murray-Rust, M. L. Lupetti, I. Nicenboim, and W. V. D. Hoog, “Grasping AI: experiential exercises for designers,” AI & Soc, vol. 39, no. 6, pp. 2891–2911, Dec. 2024, doi: 10.1007/s00146-023-01794-y.

[36] G. Majgaard, “A Pilot Study: Engineering Students use Generative AI to Support the Development of Playful Educational Technology,” 2024.

[37] A. W. W. Eide, “Applying generative adversarial networks for anomaly detection in hyperspectral remote sensing imagery,” Master’s Thesis, NTNU, 2018.

[38] M. Hosseini and K. Holmes, “The Evolution of Library Workplaces and Workflows via Generative AI,” CRL, vol. 84, no. 6, 2023, doi: 10.5860/crl.84.6.836.

[39] R. J. Krumsvik, “Artificial Intelligence, Education and the Professional Perspective,” NJDL, vol. 19, no. 2, pp. 55–63, Jun. 2024, doi: 10.18261/njdl.19.2.1.
[40] S. W. Beck and S. R. Levine, “Backtalk: ChatGPT: A powerful technology tool for writing instruction,” Phi Delta Kappan, vol. 105, no. 1, pp. 66–67, Sep. 2023, doi: 10.1177/00317217231197487.
[41] E. Bittner, S. Oeste-Reiß, and J. M. Leimeister, “Where is the Bot in our Team? Toward a Taxonomy of Design Option Combinations for Conversational Agents in Collaborative Work,” presented at the Hawaii International Conference on System Sciences, 2019. doi: 10.24251/HICSS.2019.035.
[42] K. North, “Learning in the Year 2030,” in Managing Work in the Digital Economy, S. Güldenberg, E. Ernst, and K. North, Eds., in Future of Business and Finance. , Cham: Springer International Publishing, 2021, pp. 223–238. doi: 10.1007/978-3-030-65173-2_14.
[43] F. Sandelin, “A multi-case study on generative AI use in work and decision-making: exploring applications and potential departmental effects,” Masterthesis, Svenska handelshögskolan, 2024.
[44] A. Rosignoli, R. Geyer, and S. Lenka, “The Influence of Generative AI on Creativity in the Front End of Innovation,” 2024.
[45] V. M. Danry, “AI enhanced reasoning: Augmenting human critical thinking with AI systems,” PhD Thesis, Massachusetts Institute of Technology, 2023.
[46] R. Ribeiro, “Conversational Generative AI Interface Design: Exploration of a hybrid Graphical User Interface and Conversational User Interface for interaction with ChatGPT,” University of Malmö, 2024.
[47] D. H. Zhu and Z. Z. Deng, “Effect of Social Anxiety on the Adoption of Robotic Training Partner,” Cyberpsychology, Behavior, and Social Networking, vol. 24, no. 5, pp. 343–348, May 2021, doi: 10.1089/cyber.2020.0179.
[48] T. Herrmann, “Evolution of interaction-free usage in the wake of AI,” i-com, vol. 0, no. 0, Jun. 2024, doi: 10.1515/icom-2024-0005.
[49] P. Ritala, M. Ruokonen, and L. Ramaul, “Transforming boundaries: how does ChatGPT change knowledge work?,” JBS, vol. 45, no. 3, pp. 214–220, Mar. 2024, doi: 10.1108/JBS-05-2023-0094.
[50] T. Herrmann, “Design issues for supporting collaborative creativity,” in Proc. of the 8th Int. Conf. on the Design of Cooperative Systems, 2008, pp. 179–192.
[51] A. Carell, T. Herrmann, A. Kienle, and N. Menold, “Improving the coordination of collaborative learning with process models,” in Proceedings of th 2005 conference on Computer support for collaborative learning: learning 2005: the next 10 years!, International Society of the Learning Sciences, 2005, pp. 18–27.
[52] M. J. Cubas and B. H. Ersdal, “AI Literacy at The Norwegian University of Technology and Science: A Mixed Methods Approach Measuring AI Literacy Among Technology Students,” Master’s Thesis, NTNU, 2024.
[53] J. Magrill and B. Magrill, “Preparing Educators and Students at Higher Education Institutions for an AI-Driven World,” TLI, vol. 12, Jun. 2024, doi: 10.20343/teachlearninqu.12.16.
[54] M. Prilla, T. Herrmann, and M. Degeling, “Collaborative Reflection for Learning at the Healthcare Workplace,” in Computer-Supported Collaborative Learning at the

Workplace, Sean P. Goggins, I. Jahnke, and V. Wulf, Eds., in Computer-Supported Collaborative Learning Series, no. 14. , Springer US, 2013, pp. 139–165. Accessed: Jul. 02, 2013. [Online]. Available: http://link.springer.com/chapter/10.1007/978-1-4614-1740-8_7

[55] T. Herrmann, "Conscious AI and Human Communication," 2024, doi: 10.18420/MUC2024-MCI-WS12-382.

[56] T. Herrmann and S. Pfeiffer, "Keeping the organization in the loop: a socio-technical extension of human-centered artificial intelligence," AI&Soc, vol. 38, pp. 1523–1542, 2023, doi: 10.1007/s00146-022-01391-5.

[57] T. Herrmann, I. Jahnke, and A. Nolte, "A problem-based approach to the advancement of heuristics for socio-technical evaluation," Behaviour & Information Technology, vol. 41, no. 14, pp. 3087–3109, 2022, doi: 10.1080/0144929X.2021.1972157.

[58] T. Herrmann, "Die Bedeutung menschlicher Kommunikation für die Kooperation und für die Gestaltung computerunterstützter Gruppenarbeit," Horst Oberquelle, Kooperative Arbeit und Computerunterstützung: Stand und Perspektiven. Verlag für Angewandte Psychologie Göttingen, Verlagsgruppe Hogrefe, 1991.

[59] A. Schmidt and T. Herrmann, "Intervention user interfaces: a new interaction paradigm for automated systems," interactions, vol. 24, no. 5, pp. 40–45, Aug. 2017, doi: 10.1145/3121357.